# Modal abundances of CAIs: Implications for bulk chondrite element abundances and fractionations.

Dominik C. Hezel, Sara S. Russell, Aidan J. Ross, Anton T. Kearsley

Impacts and Astromaterials Centre (IARC), Natural History Museum, Department of Mineralogy, Cromwell Road, SW7 5BD, London, UK

*E-mail address of the corresponding author: d.hezel@nhm.ac.uk

Operating system: Mac OS X 10.4.11
Word processor: Apple Pages
Spelling: British English

**Abstract**

Modal abundances of Ca,Al-rich inclusions (CAIs) are poorly known and reported data scatter across large ranges. We combine reported CAI modal abundances and our own set, and present a complete list of CAI modal abundances in carbonaceous chondrites. This includes (in area%): CV: 2.98, CM: 1.21, Acfer 094: 1.12, CO: 0.99, CK/CV (Ningqiang & DaG 055): 0.77, CK: 0.2, CR: 0.12 and CB: 0.1. CAIs are Poisson distributed and if only small areas (<1000 $mm^2$) are studied, the data are probably not representative of the true CAI modal abundances, explaining their reported large scatter in a single chondrite group. Carbonaceous chondrites have excess bulk Al concentrations when compared to the CI-chondritic value. We find a correlation between this excess and CAI modal abundances and conclude that the excess Al was delivered by CAIs. The excess Al is only a minor fraction (usually ~10 rel%, but 25 rel% in case of CVs) of the bulk chondrite Al and cannot have contributed much $^{26}Al$ to heat the chondrite parent body. Ordinary, enstatite, R- and K-chondrites have an Al deficit relative to CI chondrites and only very low CAI modal abundances, if any are present at all. Carbonaceous chondrites also had an initial Al deficit if the contribution of Al delivered by CAIs is subtracted. Therefore all chondrites probably lost a refractory rich high-T component. Only minor amounts of CAIs are present in the matrix or have been present in the chondrule precursor aggregates. Most CAI size distributions contain more than one size population, indicating that CAIs from within a single meteorite group had different origins.

## 1. Introduction

Chondrites consist primarily of two major components, chondrules and matrix and three minor components, metal, sulfides and Ca,Al-rich inclusions (CAIs). The modal abundances of the major components are well known, in contrast to the modal abundances of the minor components, which are poorly known. There exists no compilation of CAI modal abundances in chondrites in the literature, although this is an important parameter for several problems in cosmochemistry, e.g. bulk chondrite compositions, as we will show later. CAI modal abundances are often grouped together with amoeboid olivine aggregates (AOAs; e.g. Scott & Krot, 2006; Table 1), i.e. CAI modal abundances are hidden in the displayed numbers. There are several original papers and abstracts containing CAI modal abundances that we compiled in Table 2, however, the values often spread over a large range for a single chondrite group (Fig. 1; McSween 1977a, McSween 1977b, McSween 1979, Simon & Haggerty, 1979, Kornacki & Wood 1984, Rubin et al. 1985, Kallemeyn et al. 1991, Noguchi 1993, Scott et al. 1996, Russell et al. 1998, Rubin 1998, May et al. 1999, Aleon et al. 2002, Krot et al. 2002, Norton & McSween 2007). We show that this spread is the result of a Poisson distribution of the CAIs within the chondrites. A Poisson distribution represents the spatial distribution of a small number of particles that is randomly mixed with a larger number of particles, as is the case for CAIs within a chondrite. A characteristic feature of Poisson distributions is that some areas of the chondrite contain only few and others many CAIs. This feature is more pronounced with smaller areas studied and vanishes with larger areas. Here we explore the size of the area that has to be studied in order to obtain small errors for the measured CAI modal abundances. We provide a new set of CAI modal abundances that we obtained for all carbonaceous chondrites except CH and CI chondrites.

Beside this 'classical' approach of simply counting the number of CAIs in a chondrite, it is also possible to make a few assumptions and theoretically calculate their modal abundances from bulk chondrite element-concentrations or element-concentrations of individual chondrite components. CAIs are dominated by refractory elements, such as Ca and Al. The Ca/Al ratios of chondrites is close to CI-chondritic in all chondrites (e.g. Lodders 2003) and we arbitrarily choose Al to calculate CAI modal abundances, i.e. we assume Al to be representative of the refractory element abundances in CAIs. In the first approach it is assumed that all chondrites start with the same, i.e. CI-chondritic bulk chondrite Al concentration. Higher than CI-chondritic Al abundances in a chondrite are attributed to the addition of CAIs and their modal abundance can be calculated from the excess of bulk chondrite Al when compared to the CI-chondritic value. In case of lower than CI-chondritic bulk chondrite Al abundances refractory rich high-T components have probably been lost. In the second approach, a bulk chondrite Al concentration without CAIs is calculated from the Al abundances in chondrules and matrix. This Al can be subtracted from the measured bulk chondrite Al. The difference must be made up by CAIs and their modal abundance can be calculated. It has to be taken into account that some CAIs might have been part of the chondrule precursor material and might also be present in the fine-grained matrix. This second calculation is less accurate as some of the required parameters are only poorly known. The results of both calculations and also the result from counting CAIs in chondrites, and considering their Poisson distribution are in very good agreement.

There are many open question regarding the origin and formation of CAIs (e.g. MacPherson et al., 2005). We use the new CAI modal abundances we present here to conclude that (i) CAIs contributed little to the bulk chondrite refractory element abundances, (ii) CAIs formed in a separate nebular region from chondrules and matrix and (iii) CAIs cannot have contributed a significant amount of $^{26}Al$ to heat their parent bodies.

There is some confusion in the literature about the usage of vol% and area% when giving modal abundances. Most authors use vol% although they measure areas. In this study modal abundances are reported as area%. However, some calculations involving the addition or subtraction of CAIs make use of vol% because the use of area% does not make sense in this case. However, the results are reported as area%, because the mixed usage of vol% and area%, even when appropriate, would probably cause too much confusion.

## 2. Technique

### *2.1. Modelling*

Ca,Al-rich inclusions are a minor component in chondritic meteorites. If they were randomly mixed into the chondrite parent body their occurrence follows a Poisson distribution with the probability density function (pdf)

$$f(x) = \frac{\lambda^x}{x!} e^{-\lambda} \quad (1).$$

The parameter λ defines the shape of the function and represents both the mean value and the variance of the distribution, i.e. λ has the same value as the modal abundance of CAIs. CAI modal abundances are usually obtained from chondrite thin sections. We developed a model using *Mathematica 5.1* that simulates a random distribution of CAIs in an area of 100x100 mm (Fig. 2a). The area is divided into quadratic cells of identical edge lengths (Fig. 2b). The modal abundance of CAIs in each of these cells is counted and then plotted as a histogram. Inputs to the model are: (1) the size of the area studied; (2) the number of CAIs within this area; (3) assumptions that the radii of the CAIs are distributed log-normally:

$$f(x) = \begin{cases} \frac{1}{\sqrt{2\pi}\sigma x} \exp\left(-\frac{(\ln x-\mu^2)}{2\sigma^2}\right) & x > 0 \\ 0 & x \leq 0 \end{cases} \quad (2)$$

with $\mu$ and $\sigma$ defining the shape of the distribution. These two parameters and the maximum CAI radius must be specified; (4) the grid spacing into which the 100x100 mm area is divided (e.g. 10 mm in Fig. 2b).

Outputs of the model are: (1) the mean value of the CAI modal abundance (= the true modal abundance of CAIs); (2) the mean value of CAI radii; (3) the histogram with the modal abundance of CAIs on the x-axis and the number of cells with a certain modal abundance of CAIs on the y-axis. In addition graphical outputs of the CAI distribution can be produced (cf. Fig. 2). Poisson distributions are calculated from the histograms produced with the model using the CAI modal abundance for the parameter $\lambda$ in equation (1).

### *2.2. Modal abundance measurements*

We obtained CAI abundance data for CV, CR, CO, CK, CM and CB chondrites and the ungrouped chondrite Acfer 094. We used false coloured X-ray maps of thin sections to identify a total of 2049 potential CAIs. The X-ray maps were produced by combining Al (white), Ca (yellow), Mg (green), Si (blue) and Fe (red). An example of the Allende 1 sample is displayed in the electronic appendix. All objects that appeared to have high Al and/or Ca concentrations in the X-ray map were considered to be CAIs. We crosschecked a part of the identified CAIs with the electron microscope and found that about 95% of the CAIs identified using the X-ray maps are CAIs, i.e. are not mesostasis fragments, the only other candidate we consider to have high Al and/or Ca. We considered everything to be mesostasis that has a close to feldspathic composition. From this, we estimate the total error of this technique to be <5%, which is very small considering the low modal abundances of CAIs. A modal abundance of 3 area% would have an absolute error of only ±0.15 area%.

There are different groups of CAIs (e.g. Brearley and Jones, 1998). We did not classify them, as this would be beyond the scope of this paper. The smallest CAI that can be identified on an X-ray map is about half the size of a single pixel on the map, because the Ca and/or Al of the CAI still contributes enough X-rays to make this a distinct pixel representing a high Ca and/or Al spot. The edge lengths of a single pixel in the different maps we used are listed in Table 3. Pixel sizes vary depending on sample size. In order to obtain element maps within a reasonable time, larger samples have larger pixel sizes. Most CAIs we measured are larger than 10 and nearly all are larger than 5 pixels. As there are nearly no CAIs smaller than 5 pixels and as these are quite small, CAIs smaller than 5 pixels contribute only marginally to the total CAI modal abundance.

CAIs often have irregular outlines. In order to provide an intuitively understandable size of the CAIs, these are assumed to be circles and 'model radii' $r_m$ are calculated using the CAI areas ($A_{CAI}$), which are easy to calculate from the image processed data. The model radius is then calculated as follows

$$r_m = \sqrt{\frac{A_{CAI}}{\pi}} \qquad (3).$$

## 3. Results

### *3.1. Modelling*

Figure 2 shows how a random distribution of 3500 CAIs leads to regions with high and low CAI densities. The modal abundance of CAIs in this figure is 2.61 area%, the average CAI radius is 113 µm and the maximum CAI radius is 500 µm. Figure 3 displays histograms obtained from Fig. 2. Individual histograms correspond to different cell edge lengths of the grid displayed in Fig. 2b. If the sample areas are small, e.g. the edge length of a single cell is only 5 mm, only about 23% of all cells contain the true CAI modal abundance of 2.61 area%. About 42% of the cells contain lower and 35% higher than the true modal abundance. Increasing the sample size, i.e. the edge length of a single cell, narrows this distribution. After reaching a sample size of 2500 $mm^2$ (= 50 mm edge length of a single cell), all of the cells are representative of the true CAI modal abundance. Note that bin ranges of the histograms in Fig. 3 have been chosen large (1 area%) in order to keep the plot readable. The 3 area% bin of the histogram with 20 mm edge length of a single cell contains many cells with modal abundances <3 area%, compensating for some of the cells in the bin of 4 area%. An average chondrite thin section sample might have an area of 100 $mm^2$, equal to a square with an edge length of 10 mm. In this case, as can be seen from Fig. 3 only 40% of all cells are representative of the true CAI modal abundance. The rest mainly give CAI abundances that are too high. The range of CAI modal abundances that can be found using a sample size of 100 $mm^2$ and the conditions defined above spans from nearly 0 to up to 10 area%. The Poisson distribution plotted into Fig. 3 (dashed grey line) is fitted to the histogram of 100 $mm^2$ samples, illustrating the detailed distribution of the latter. The histograms and the Poisson pdf shown in Fig. 3 will narrow with smaller and broaden with larger CAI modal abundances. Also the CAI sizes have an influence on the shape. Larger CAIs require even larger samples to give representative results, whereas for smaller CAIs smaller samples sizes are sufficient. As we will show, the large ranges seen in the histograms correlate with the large range of CAI modal abundances reported within an individual chondrite group.

The modelling also allows us to calculate the errors of the CAI modal abundance measurements associated with the studied sample size. The errors are calculated for a confidence interval of 95.4%, corresponding to $2\sigma$. The calculation procedure is illustrated using Fig. 3. As can be seen in this figure, there is a less then 5% chance that on a sample area of 100 $mm^2$ a CAI modal abundance with less than 1 area% is measured. This in turn means, if an area of 100 $mm^2$ is studied and 1 area% CAIs is measured, there is a 95% chance that the true CAI modal abundance is below 2.61 area%. The upper error would therefore be 1+1.61 area%. The same has to be done for the lower limit. The errors provided with our CAI modal abundance have been calculated in this way. It is noted that these errors depend on the assumption made in the model and that in order to perform this procedure in a reasonable computing time, the values were estimated as soon as these have been close to the 5% mark. The combined error from measuring the CAI modal abundance and using a certain sample size is calculated using the error propagation formula

$$\delta z = \sqrt{\delta x^2 + \delta y^2} \quad (4)$$

with $\delta z$: combined error and $\delta x$, $\delta y$: individual errors. However, the contribution of the error from measuring modal abundances is negligible.

### *3.2. Modal abundances of CAIs*

In addition to the CAI modal abundances we measured for carbonaceous chondrites and the ungrouped chondrite Acfer 094, we compiled all CAI modal abundances reported in the literature (Table 2). In the following we focus only on modal abundances of CAIs (Table 4) and their size distributions in various chondrites. Size distribution histograms are given for two different bin ranges to illustrate the effects of choosing different bin ranges (Fig. 4). We do not discuss the petrographic or petrologic appearance of CAIs. A good documentation of these can be found in Brearley & Jones (1998).

#### **3.2.1 CV**

We studied three thin sections of Allende, covering a total area of 630 $mm^2$ and found 5.06 area% of CAIs. The size distribution displayed in Fig. 4 has a pronounced peak at the smallest model radii of <50 μm. The inset in Fig. 4 displays the size distribution of model radii >150 μm. The number of CAIs decreases monotonically with increasing model radius. Previous studies show a large spread in CV CAI modal abundances (Table 2, Fig. 1). McSween (1977b) reported CAI modal abundances ranging from 2.5 to 9.4 area% and May et al. (1999) a much narrower range between 0.65 and 1.89 area%. These ranges represent different CV chondrites, but even data for only Allende spread over a wide range: Kornacki & Wood (1984) reported 2.52 area%, we found 5.06 area%, Simon & Haggerty (1979) reported 7.1 area% (this is a recalculated value: Simon & Haggerty (1979) report a value of 15%, but normalised to the total of all large components, neglecting matrix. If a total abundance of large components of 45% is assumed, the reported 15% reduces to 7.1%) and McSween (1977b) report 9.4 area%.

### 3.2.2. CK & CK/CV-like

We analysed three different CKs: one thin section of Karoonda (146 $mm^2$), two of Ningqiang (total: 185 $mm^2$) and one of DaG 055 (169 $mm^2$). However, only Karoonda is officially classified as a CK, the other two are CV/CK-like and it is recommended to designate them as ungrouped. Nonetheless, Greenwood et al. (2004) pointed out certain affiliations of Ningqiang and DaG 055 with the CKs. We found no CAIs in Karoonda, 0.94 area% in Ningqiang and 0.59 area% in DaG 055. The CAI size distributions of Ningqiang and DaG 055 resemble log-normal distributions (Fig. 4). From the data reported by Noguchi (1993) for Karoonda and EET 87507 we calculate a CAI modal abundance of 0.2 area%. Kallemeyn et al. (1991) reported 1.0 area% of CAIs in Ningqiang, which is in agreement with our data.

### 3.2.3. CM

We analysed one thin section of Murchison and one of Nogoya; both have quite small areas (34.4 and 6.2 $mm^2$, respectively). We found 0.97 area% CAIs in Murchison and 0.016 area% CAIs in Nogoya. The CAI sizes do not follow a simple distribution, which is probably due to the small areas analysed. Norton & McSween (2007) report an average CAI abundance of 1.6 area%, considerably lower than the 4.8 area% reported for Murray by McSween (1997b).

### 3.2.4. CO

We studied CAI abundances in three different CO chondrites: one thin section of Warrenton (42 $mm^2$, containing 1.01 area% CAIs), one of Kainsaz (142 $mm^2$, containing 0.47 area% CAIs) and one of DaG 190 (197 $mm^2$, containing 1.35 area% CAIs). The combined size distribution of all three chondrites resembles a log-normal distribution, with a few CAIs of larger sizes (>330 μm). COs are the second group whose CAI modal abundances have been previously extensively studied, the other being CVs. Like the CVs, the range of reported CAI modal abundances is quite large (Fig. 1), although not as large as in CVs, which is probably because the true CO CAI modal abundance is smaller. Rubin et al. (1985) reported a range of 1.0 to 3.6 area% for three samples of Colony and 1.2 area% CAIs for ALH 77307. McSween (1977a) studied six different COs with reported values between 1.2 and 3.5 area%. Russell et al. (1998) studied ten different CO chondrites and found a range of CAI abundances between 0.63 and 1.5 area%. Finally, May et al. (1999) list CAI abundances for three different COs, ranging from 0.85 to 1.38 area%.

### 3.2.5. CR

We studied one thin section of Renazzo (112 $mm^2$) and one of Acfer 209 (155 $mm^2$) and found in both extremely low CAI abundances. In Renazzo we measured 0.09 area% and in Acfer 209 0.14 area%. The CAI size distribution can be regarded as log-normal with a few CAIs at large sizes (>140 µm). Previously, only McSween (1977b) reported CAI modal abundances for Renazzo (0.3 area%) and Al Rais (0.8 area%), which are similarly low.

### 3.2.6. CB

We studied two thin sections of Isheyevo with a combined area of 335 $mm^2$. The CAI abundance is extremely low, about only 0.01 area%. The CAI size distribution is log-normal-like although displaying some gaps. We found no further quantitative reports on CAI abundances in CB chondrites.

### 3.2.7. Acfer 094

We studied one small thin section of the ungrouped chondrite Acfer 094 (21.97 $mm^2$) and found an abundance of 1.12 area% CAIs. This is agreement with the only other report on CAIs in Acfer 094 with <2 area% by Weber (1995). The size distribution of the CAIs we studied resembles a log-normal distribution with a few gaps.

### 3.2.8. Other chondrites

We did not extend our study to ordinary chondrites, enstatite chondrites, CI-, R- or K-chondrites. Their reported modal abundances of refractory inclusions are extremely low, usually below 0.1 area% (Table 1). It would be necessary to study extremely large areas to obtain accurate numbers.

### *3.3. Theoretical calculation of CAI modal abundances*

#### 3.3.1. Using bulk chondrite Al concentrations

The Si-normalised Al abundances of CI-chondrites and the solar photosphere are in excellent agreement (1%; e.g. Lodders, 2003). It is therefore assumed here that Al was homogeneously distributed in the solar system and that deviations from the chondritic Al concentration were established shortly before or during chondrule formation. One such process was probably the addition or subtraction of high temperature components such as CAIs. The amount of Al that was either added or subtracted can be represented by the difference between bulk CI-chondritic Al concentration and the bulk Al concentration of different chondrite groups. Element concentrations can only be given in relative amounts, usually in wt%. If the amount of any element changes, the relative concentrations of all other elements also change. A direct comparison of the Al concentrations of different chondrite groups and the CI-chondritic Al concentration is therefore not possible. In order to compare increases or decreases of Al in the different chondrite groups relative to the CI-value, we re-calculate their bulk Al concentrations and obtain new values that represent bulk Al concentrations of the chondrites as if these had not been changed from the CI-chondritic composition.

Element ratios do not change if the amount of any element other than those used in the ratio changes. An element ratio can be written as Al/x, with x being any element. If x is an element in the chondrite ($x_{chondrite}$) whose amount has not changed relative to its CI-chondritic value, the $Al_{CI}/x_{CI}$ ratio equals the $Al_u/x_{chondrite}$ ratio. $Al_u$ is the sought-after, unchanged (*u*) Al concentration that can be easily calculated using the formula

$$Al_u = x_{chondrite} \cdot \frac{Al_{CI}}{x_{CI}} \qquad (5).$$

The crucial assumption is that the amount of element $x$ in the chondrite has not changed relative to the CI-chondritic amount. It is therefore necessary to choose an appropriate element for which this assumption is valid. The main elements Mg, Si and Fe (together making up 80-90 wt% of chondrites if oxygen is neglected) should be the least affected by fractionation processes. This is particulary evident in carbonaceous and K-chondrites, in which the Mg/Si ratios are the same as in CI-chondrites. All other chondrites have fractionated Mg/Si ratios. The implications if one of these elements was fractionated are discussed below. Because fractionations are, however, possible in one or more of these elements we chose to substitute $x_{chondrite}$ with all three elements, thereby obtaining a range of CAI modal abundances. Columns 1-4 of Table 5 list the bulk chondrite element concentrations used for the calculations (taken from Lodders and Fegley, 1998 and Zipfel et al. 1998) and columns 5-7 display the $Al_u$ concentrations when $x_{chondrite}$ (designated simply as $x$ in Table 5) is substituted by Mg, Si and Fe. Columns 8-10 list the difference of the measured bulk chondrite Al concentration $Al_{chondrite}$ and $Al_u$. This number represents an excess or deficit of Al in a particular chondrite relative to the CI-chondritic value.

It is then assumed that all excess Al was delivered by CAIs. The amount of CAIs required to account for this excess Al is calculated using the formula

$$MA_{CAI} = \frac{\Delta Al \cdot 100}{Al_{CAI}} \qquad (6)$$

with $MA_{CAI}$: modal abundance of CAIs (calculated in vol%, but given in area%, cf. introductory remarks) and $Al_{CAI}$: bulk Al concentration of CAIs. We used an $Al_{CAI}$ of 18 wt%. Simon and Grossman (2004) measured the chemical bulk compositions of 6 Type A and 17 Type B CAIs. Their analysed Type A CAIs have Al concentrations between 17.05 and 20.72 wt% and an average of 18.63 wt%, and their Type B CAIs have Al concentrations between 16.09 and 18.91 wt% with an average of 17.35 wt%. An $Al_{CAI}$ of 18 wt% therefore seems to be a reasonable average. Higher values for $Al_{CAI}$ will result in lower CAI modal abundances and conversely lower values for $Al_{CAI}$ will result in higher CAI modal abundances. The results of the calculations are displayed in columns 11-13 of Table 5, together with our newly determined CAI modal abundances for chondrites (column 14, cf. Table 6). The calculated values using $x_{chondrite}$ = Mg and $x_{chondrite}$ = Si are similar. This is not surprising, because all carbonaceous chondrites have virtually the same Mg/Si ratios. Iron has a much larger variation among all chondrites than Mg or Si. The differences for $x_{chondrite}$ = Mg, $x_{chondrite}$ = Si and $x_{chondrite}$ = Fe vary between a factor of 1.13 and 2.37. Iron is probably more affected by fractionation processes, either through redox reactions or because it is more volatile than Mg or Si. We therefore choose to use the CAI modal abundances calculated using $x_{chondrite}$ = Mg for the discussions below.

Finally, another column (15) contains CAI modal abundances calculated under the assumption that all Al in the chondrites is concentrated in CAIs. This value provides the absolute maximum modal abundance of CAIs. Again a bulk CAI Al concentration of 18 wt% is used for this calculation. The result is indeed a highly theoretical number, but provides an estimate of maximum CAI modal abundances and what could be expected if all Al found in chondrules and matrix was initially delivered by CAIs. CV chondrites have the largest excess of Al relative to the CI-chondritic value and if all Al were delivered by CAIs, they could not contain more than 9.3 area% CAIs; this is therefore the theoretical absolute maximum CAI modal abundance for all chondrites.

### 3.3.2. Using Al concentrations of individual chondrite components

A second way to theoretically determine CAI modal abundances is to calculate a 'reduced' bulk chondrite Al concentration from the Al concentrations of all meteoritic components like chondrules and matrix, but without CAIs (therefore the designation 'reduced'). The difference between this reduced calculated and the measured bulk chondrite Al concentration is used in the same way as above to calculate a theoretical CAI modal abundance. However, the available data set of bulk chondrule and matrix Al concentrations as well as the required modal abundances of chondrules, matrix, metal and sulfides is insufficient and often contradictory and unfortunately makes this alternative way of theoretically calculating accurate CAI modal abundances very difficult. However, this approach is still feasible for a reasonable estimate for the CV chondrites, where some data are available.

Chondrules and matrix contain a significant amount of Al that cannot be attributed to a CAI origin. Most chondrules have flat to near flat REE element patterns (e.g. Hezel et al. 2006, Pack pers. com.), but CAIs often have largely fractionated REE patterns (e.g. MacPherson et al. 1988) and carry a significant amount of REE. If Al were entirely inherited from CAIs, REEs would also be more or less entirely inherited from CAIs and their fractionated patterns would show up in the bulk chondrule compositions. This effect can be used to identify the addition of CAIs to chondrules (Pack et al. 2004, MacPherson and Huss 2005, Hezel et al. 2006, Hezel and Palme 2007). As most chondrules have unfractionated REE patterns and comparatively minor REE enrichments, no more than 1-5 vol% CAIs were added to a single chondrule, depending on the assumptions made for REE element concentrations in CAI-free chondrules and added CAIs. Chondrules in CV chondrites have reported bulk Al concentrations between 2.17 and 2.80 wt% (Rubin and Wasson 1987,

Kimura and Ikeda 1998). If we assume that approximately 1-3 vol% CAIs have been added to chondrules, about 1.8 wt% Al in bulk chondrules is of non-CAI origin. Reported matrix analyses of CV chondrites span a range from 1.03-1.62 wt% (Clarke et al. 1970, McSween and Richardson 1977, Rubin 1984, Rubin and Wasson 1987, Jarosewich 1990, Kimura and Ikeda, 1998, Klerner 2001, Hezel and Palme 2008,). Matrix and chondrules are most probably genetically linked as they show a complementary relationship with respect to certain element ratios (Mg/Si - Klerner and Palme, 1999; Ti/Al - Klerner and Palme 2000; Ca/Al - Hezel and Palme, 2008) and have therefore most probably formed in a chemically closed nebula environment. This genetic relationship and the fact that most Al in chondrules is not from CAIs makes it equally unlikely that most Al in the matrix is of CAI origin. If we assume again that approximately 1-3 vol% CAIs have been added to the matrix 0.8 wt% Al is of non-CAI origin. Assuming a metal and/or sulfide abundance of 4 area% and a chondrule and matrix abundance each of 48 area% results in a reduced CV bulk chondrite Al concentration of 1.17 wt% Al. The difference between the measured bulk CV chondrite Al concentration and this estimated reduced bulk composition is 0.51 wt% Al. If this additional Al in CV chondrites was delivered by CAIs with a bulk Al concentration of 18 wt%, 3.0 area% of CAIs are required. This rough estimate is identical to the modal abundance we calculated in the first approach as well as to our deduced value listed in Table 6. Although this is a nice result, it is again noted that bulk chondrule and matrix Al concentrations have comparatively large errors and these results need to be treated with caution until more accurate data are available to perform this calculation reliably. In fact, as we will point out below, the contribution of CAIs to either chondrules or matrix was most probably below 1 vol%. From this it seems the reported Al concentrations for chondrules and matrix in the literature are probably too high. The sole purpose of this calculation here is to demonstrate that the present data base of chondrule and matrix Al abundances readily allow rough estimates of CAI modal abundances.

Theoretical CAI modal abundances should be lower in other chondrites, as these have lower than CV bulk Al concentrations, but probably similar Al concentrations in their chondrules and matrices.

## 4. Discussion

### *4.1. CAI modal abundance measurements*

Here we present the first compilation of CAI modal abundances in chondrites (Table 6). The reported modal abundances of individual chondrite groups often scatter over a large range. This scatter is produced by the Poisson distribution of CAIs in chondrites. If the sample areas studied for CAI modal abundances are too small, the result is not representative of the true CAI modal abundance. The size of the area that needs to be studied depends on parameters like the approximate expected CAI modal abundance, average size of the CAIs, their size distribution etc. A reliable modal abundance, i.e. a modal abundance with only a small error, can be obtained if the sample size is in the range of at least 1000-2000 $mm^2$. Smaller sizes probably result in an underestimate of the true modal abundance (Fig. 3). To achieve these large sample areas it is usually necessary to study multiple samples. However, it is not possible to calculate a simple average of CAI modal abundances obtained from such different samples. The Poisson distribution of the CAIs implies that each single area is not representative of the true CAI modal abundance. A weighted average must be calculated, i.e., all measured CAI areas from all samples must be summed up and all sample areas must be summed up and divided by each other to obtain the correct weighted average $av_{wght}$:

$$av_{wght} = \frac{\sum area_{CAIs}}{\sum area_{samples}} \quad (7).$$

The area studied when reporting CAI modal abundances is essential information that must always be given, as it is an indicator of how large the error is that accompanies the measurement. We produced a table of CAI modal abundances in chondrites (Table 6). The values in this table are weighted averages of our newly obtained data and all data reported in the literature that also list their studied sample sizes. Details on the literature used are given in the caption of Table 2. Only the bold data are new, the others are taken from Table 1 and might be regarded as upper limits. This is feasible as virtually all of them are <0.1 area%. It is also clear that the maximum error when determining CAI modal abundances equals the true CAI modal abundance of the chondrite (as long as the studied sample size is reasonably large). That is, small modal abundances result in very small errors and in many cases it might not be necessary to determine CAI abundances <1 area% with high accuracy.

## *4.2. Refractory element contents and fractionations in chondrites*

### 4.2.1. Ordinary, enstatite R- and K-chondrites

Virtually all ordinary, enstatite, R- and K-chondrites have Al deficits, between -0.07 and -0.63 wt% (Table 5, columns 11-13) and in most cases no CAI modal abundances can be calculated for them. This is in agreement with the extremely rare occurrence of CAIs in these chondrites. We suggest that a high-T condensate carried away a significant quantity of refractory elements prior to parent body formation. Ordinary, enstatite and R-chondrites also all have lower than CI-chondritic Mg/Si ratios. If their initial Mg/Si ratios were the same as the CI-chondritic value, Mg had to be effectively removed or Si to be effectively added in order to achieve the lower than CI-chondritic Mg/Si ratios. It has often been suggested that forsterite, a high-T component, was removed from these chondrite forming regions. As a consequence, the Al/Mg ratios must increase. However, the Al/Mg ratios of ordinary, enstatite and K-chondrites are lower than the CI-chondritic value. As the loss of Mg increases the Al/Mg ratios, the regions where ordinary, enstatite and K-chondrites formed must have lost even more Al than is currently recorded in their sub-CI-chondritic Al/Mg ratios, i.e. even more of the high-T refractory phase must have been lost. Alternatively, in order to simultaneously decrease the Al/Mg ratio and increase the Mg/Si ratio it is also possible to add low-T material with a bulk Mg/Si ratio $\ll$1 (e.g. feldspars or Mg-poor phyllosilicates). In this case the addition of Mg can account for the lower than CI-chondritic Al/Mg ratios in OCs, ECs and K-chondrites. The available data are insufficient to unequivocally decide whether the removal of a component with Mg/Si > 1 or the addition of a component with Mg/Si $\ll$ 1 is responsible for the observed ratios, although the removal of forsterite seems more plausible, with the consequence that even more refractory elements must have been lost, too, in order to reduce the Al/Mg ratio. Addition of Si could account for the lower Mg/Si ratio, but would not affect the Al/Mg ratio.

### 4.2.2. Carbonaceous chondrites

Columns 11-13 of Table 5 show that carbonaceous chondrites, except for CRs, have Al excesses between 0.02 and 0.56 wt%. Figure 5 is a plot of calculated versus measured CAI modal abundances. Only chondrites with an Al excess relative to the CI-chondritic Al concentration are plotted. The Al excesses in this plot were calculated assuming that Mg did not change relative to its CI-chondritic abundance. The dashed line represents the amount of CAIs calculated from the Al excess and is therefore designated the ‘line of maximum CAI addition’. If the total Al excess is the result of CAI addition, their modal abundance must plot on the line of maximum CAI addition. If CAI modal abundances fall below this line, the Al excess cannot be explained by CAI addition alone. If CAI modal abundances fall above the line, the chondrites must have had an Al deficit prior to CAI addition, as the CAIs deliver more Al than available from the calculated Al excess. CM, CO and CV chondrites plot slightly above the line, CH chondrites plot slightly below, but very close to the line and CK chondrites plot far below the line. Two horizontal error bars represent different bulk CAI Al concentrations, ranging from 14 to 22 wt%. CAIs should not have much less than 14 wt% Al as they are dominated by high-Al phases such as (numbers are Al concentrations in wt%): corundum: 53; grossite: 52; hibonite: 40; spinel: 38; gehlenite: 20; anorthite: 19; fassaite: 12 and Al-diopside: 7. As can be seen from Fig. 5, even low bulk CAI Al concentrations such as 14 wt% do not push the CAI modal abundances of CO, CM and CV chondrites much closer to the line of maximum CAI addition. Vertical bars represent the errors as determined in section 3.1. Although these error bars are quite large, CM, CO, CK and CV chondrites still fall beyond the line of maximum CAI addition. If the Al contributions of CAIs are subtracted from CM, CO, CK and CV chondrites, their remaining bulk Al concentration must be smaller compared to the CI-chondritic Al concentration. It is therefore possible that these carbonaceous chondrites also lost Al like the ordinary, enstatite, K- and R-chondrites probably also carried away by some high-T condensate rich in refractory elements. This loss was later compensated, in fact over-compensated, by the addition of CAIs.

In contrast to the CM, CO and CV-chondrites, if the Al contribution of CAIs is subtracted from CK chondrites, these have a higher than CI-chondritic bulk Al concentration. The expected CAI modal abundances of CKs calculated from their bulk Al concentration is 0.88 area% CAIs (Table 5, column 11). In contrast, our suggested observed CAI modal abundance is only 0.2 area% (Table 6). It is possible that CK chondrites are the only group that initially did not lose, but gained Al by another process than CAI addition. However, as CK chondrites are usually thermally metamorphosed, it seems more likely that CKs initially contained more CAIs that were subsequently reprocessed during heating. The CK chondrite Karoonda we analysed contains plenty of feldspar aggregates sustaining this view.

The CH and CB chondrites are the only group that have neither sub-chondritic nor higher than expected Al/Mg ratios. These chondrites also have not experienced extensive thermal metamorphism. These might therefore be the only groups with approximately bulk CI-chondritic Al concentrations, relative to an assumed, CI-chondritic Mg concentration, if the small Al contribution of CAIs is subtracted.

We note again that this discussion relies on the assumption that the bulk chondrite Mg abundance is the same as the CI-chondritic Mg abundance. However, if a change in the Mg concentration was responsible for the change in the Al/Mg ratio, it is required that large amounts of Mg were either added or subtracted. In case of carbonaceous and K-chondrites it would be required that the same amounts of Si were added or subtracted, as the Mg/Si ratios of all carbonaceous and K- chondrites are similar to the CI-chondritic Mg/Si ratio. An addition or subtraction of Mg seems therefore highly improbable and the only other way to change the Al/Mg ratio is to add or subtract Al, as described above.

### *4.3. Implications for the origin of CAIs and parent body heating*

It was recognised early by Larimer (1979) that CAIs, carbonaceous chondrites and CI chondrites plot along a mixing line in the Mg/Al vs. Si/Al diagram with CAIs and CI chondrites being the end members of this line. Larimer (1979) noticed that the mixing line might look as if the addition of CAIs explains the excess of Al in carbonaceous chondrites relative to CI, however, he argued that CAIs have non-CI-chondritic Mg/Si ratios and hence the addition of CAIs should alter the Mg/Si ratio of carbonaceous chondrites to non CI-chondritic values. However, this argument was probably based on the CAI modal abundances reported two years earlier by McSween (1977), which are partialy too high, as outlined above. In light of our measured and calculated CAI modal abundances, the concern expressed by Larimer (1979) that CAIs would disturb the carbonaceous chondrite CI-chondritic Mg/Si ratios are negligible. Klerner (2001) analysed 8 chunks of Allende using XRF and found Mg/Si ratios of 0.94, 0.95, 0.90, 0.95, 0.91, 0.90, 0.93 and 0.95, with a mean of 0.93±0.02. The uncertainty of ~2% is exactly the deviation that 3 area% of CAIs - as is the case for CVs - with Mg 6.47 wt% and Si 12.75 wt% (averages from Simon & Grosmann, 2004) would introduce to a CI-chondritic Mg/Si ratio and might therefore simply not have been detected so far. The deviation will be even smaller with fewer CAIs added, which is the case for all other chondrite groups. We are currently trying to conduct high precision bulk chondrite analyses to explore whether we are able to find systematic deviations of bulk chondrite Mg/Si ratios from the CI-chondritic value.

The relatively good match between theoretically predicted and reported CAI abundances leave little room for large quantities of CAIs hidden as tiny grains in the matrix or as precursors of chondrules. A fraction of 1 area% CAIs in the chondrule precursor assemblage and a chondrule modal abundance of 50 area% would add only 0.5 area% CAIs to the CAI modal abundance of the chondrite.

The Al excesses of carbonaceous chondrites calculated in Table 5 are around 10 rel%, except for CVs where the excess Al contributes ~25 rel%. If the excess Al is delivered by the addition of CAIs, their contribution to the bulk chondrite Al content is only minor. The major fraction of Al is contained in silicates and must be inherited from the primary chondrite source region.

Differences in O-isotopes or radiogenic $^{26}$Mg between CAIs and other chondrite components have previously been used to demonstrate a separate origin of CAIs (e.g. MacPherson et al. 2005). Our results support this and clearly show that CAIs are not a genuine component of the chemical reservoir from which other components like chondrules and matrix formed (e.g. Bland et al. 2005, Hezel and Palme 2007). CAIs were probably delivered to the chondrite forming regions during and after chondrule formation, because CAIs occur both as molten and unmolten objects. Some molten CAIs have similar radiogenic $^{26}$Mg as chondrules and were probably molten during the chondrule formation event.

The CAI size distributions of most carbonaceous chondrites more or less follow a log-normal distribution, which would most probably be the result of grinding and fragmentation during mutual collisions in the solar nebula. This is supported by their usually fragmental appearance in back scatter electron images. Some carbonaceous chondrite groups have a small population of larger CAIs, which might suggest that some CAIs in a given chondrite group stem from different source regions. This observation requires further confirmation, but is important as most studies of CAIs focus on larger specimens.

## 5. Conclusions

Our new CAI modal abundances data and our recognition that they are Poisson distributed as well as our theoretical calculations show that CAI modal abundances are much smaller than previously thought. The data reduction process is crucial when obtaining CAI modal abundances. Because CAIs are Poisson distributed, large areas must be studied (>1000, better >2000 mm$^2$) in order to obtain modal abundances with a small error. If more than one sample is studied a weighted average for the CAI modal abundance must be calculated (cf. equation 6). The CAI modal abundances we provide for carbonaceous chondrites (Table 6) are in good agreement with their calculated Al overabundance when compared to the CI-chondritic composition (Table 5). Our results support the model that CAIs did not form in the same chemical reservoir as chondrules and matrix, but were added later to this compartment.

CAIs were only occasionally added to ordinary, enstatite and R- and K-chondrites. Nearly all of these chondrites have an Al deficit compared to the CI-chondritic value. It is therefore not surprising that CAIs are extremely rare. The parental chemical reservoir of these chondrites probably lost Al by the subtraction of a refractory rich, high-T component.

Carbonaceous chondrites initially probably also had a deficit in Al and lost a refractory rich, high-T component. This loss was then over-compensated by the addition of CAIs, leading to an Al excess in carbonaceous chondrites.

Although more CAIs have been added to carbonaceous chondrites as calculated from their Al excesses, their presence among chondrule precursors as well as in the matrix is still minor (e.g. MacPherson and Huss 2005, Hezel et al. 2006, Scott 2007, Pack pers. com.). Therefore CAIs added to carbonaceous chondrites deliver only a minor fraction (usually 10 rel% and 25 rel% in case of CVs) of the carbonaceous chondrite bulk Al contents and cannot have been the source of large amounts of $^{26}$Al that substantially contributed to heat the chondrite parent body even if they contained live $^{26}$Al on accretion, as suggested by, for example the X-wind model. This is even more true for ordinary, enstatite, R- and K-chondrites, which all have Al deficits.

The CAI size distributions of nearly all CCs contain at least two different populations, of which one is small consisting of large CAIs, probably indicating that CAIs in a single chondrite group had multiple sources.

## Acknowledgements

We are grateful for thorough and constructive reviews from H. Connolly, H. Palme, J. Paque and the Associate Editor C. Floss that helped to improve and clarify the paper. This work was supported by PPARC and by a grant awarded by the Paneth Trust to A.J.R. We thank Lauren Howard for her brave assistance at the electron microprobe.

# REFERENCES

Aleon J., Krot A. N., McKeegan K. D. 2002. Calcium-aluminum-rich inclusions and amoeboid olivine aggregates from the CR carbonaceous chondrites. *Meteoritics & Planetary Science* 37:1729-1755.

Bland, P.A., Alard, O., Benedix, G.K., Kearsley, A.T., Menzies, O.N., Watt, L.E., Rogers, N.W. (2005) Volatile fractionation in the early solar system and chondrule_matrix complementarity. *Proceedings of the National Academy of Sciences 102*:13755-13760.

Brearley A. J., and Jones R. H. 1998. Chondritic meteorites. In *Reviews in mineralogy: Planetary materials,* edited by Papike J. J. Washington, D. C.: Mineralogical Society of America. 398 p.

Clark R.S., Jarosewich E., Mason B., Nelen J., Gomez M., and Hyde J.R. 1970. The Allende, Mexico, meteorite shower. *Smithsonina Contributions to Earth Sciences* 5:1–53.

Greenwood R. C., Franchi I. A., Kearsley A. T., and Alard O. 2004. The relationship between CK and CV chondrites: a single parent body source? (abstract). *35. Lunar and Planetary Science Conference* #1664.

Hezel D. C., and Palme H. 2007. The conditions of chondrule formation, Part I: Closed system. *Geochimica et Cosmochimica Acta* 71:4092-4107.

Hezel D. C., and Palme H. 2008. Constraints for chondrule formation from Ca-Al distribution in carbonaceous chondrites. *Earth & Planetary Science Letters* 265:716-725.

Hezel D. C., Palme H., Nasdala L., and Brenker F. E. 2006. Origin of $SiO_2$-rich components in ordinary chondrites. *Geochimica et Cosmochimica Acta* 70:1548-1564.

Jarosewich E., 1990. Chemical analyses of meteorites: a compilation of stony and iron meteorite analyses. *Meteoritics* 25:323–337.

Kallemeyn G. W., Rubin A. E., and Wasson J. T. 1991. The compositionally classification of chondrites: V. The Karoonda (CK) group of carbonaceous chondrites. *Geochimica et Cosmochimica Acta* 55:881-892.

Kimura M., and Ikeda Y. 1998. Hydrous and anhydrous alteration of chondrules in Kaba and Mokoia CV chondrites. *Meteoritics & Planetary Sciences* 33:1139–1146.

Klerner S. 2001. Materie im frühen Sonnensystem: Die Entstehung von Chondren, Matrix und refraktären Forsteriten. PhD-thesis, Colgone, 125pp.

Klerner S., and Palme H. 1999. Origin of chondrules and matrix in carbonaceous chondrites(abstract). *30. Lunar and Planetary Science Conference* #1272.

Klerner S., and Palme H. 2000. Large Ti/Al fractionation between chondrules and matrix in Renazzo and other carbonaceous chondrites(abstract). *63. Meteoritical Science Conference* #5131.

Kornacki A. S., and Wood J. A. 1984. Petrography and classification of Ca, Al-rich and olivine rich inclusions in the Allende CV3 chondrite. *Journal of Geophysical Research* 89:B573-B587, Supplement.

Krot A. N., Aleon J., McKeegan K. D. 2002. Mineralogy, petrography and oxygen-isotopic compositions of Ca, Al-rich inclusions and moeboid olivine aggreagates in the CR carbonaceous chondrites (abstract). *33. Lunar and Planetary Science Conference* #1412.

Lodders K. 2003. Solar system abundances and condensation temperatures of the elements. *The Astrophysical Journal* 591:1220-1247.

Lodders K., and Fegley B. jr. 1998.The Planetary Scientist's Companion. Oxford University Press, Oxford, 371p.

MacPherson G. J., and Huss G. R. 2005. Petrogenesis of Al-rich chondrules: Evidence from bulk compositions and phase equilibria. *Geochimica et Cosmochimica Acta* 69:3099-3127.

MacPherson G. J., Wark D. A., and Armstrong J. T. 1988. Primitive material surviving in chondrites: Refractory inclusion. In *Meteorites and the Early Solar System* Kerridge J. F., and Matthews M. S. eds. University of Arizona Press, 746-807.

MacPherson G. J., Simon S. B., Davis A. M., Grossman L., and Krot A. N. 2005. Calcium-Aluminum-rich inclusions: Major unanswered questions. In: *Chondrites and the Protoplanetary Disk.* Krot A. N., Scott E. R. D. and Reipurth B. (eds.), ASP Conference Series, Vol. 341, San Francisco: Astronomical Society of the Pacific, 1029p.

May C., Russell S. S., and Grady M. M. 1999. Analysis of chondrule and CAI size and abundance in CO3 and CV3 chondrites: a preliminary study (abstract). *30. Lunar and Planetary Science Conference* #1688.

McSween H. Y., Jr., 1977a. Carbonaceous chondrites of the Ornans type: a metamorphic sequence. *Geochimica et Cosmochimica Acta* 41:477-491.

McSween H. Y., Jr., 1977b. Petrographic variations among carbonaceous chondrites of the Vigarano type. *Geochimica et Cosmochimica Acta* 41:1777-1790.

McSween H. Y., Jr., 1979. Matrix variations in CM chondrites (abstract). *10. Lunar and Planetary Science Conference* #810-812.

McSween H.Y., and Richardson S.M. 1977. The composition of carbonaceous chondrite matrix. *Geochimica et Cosmochimica Acta* 41:1145–1161.

Noguchi T. 1993. Petrology and mineralogy of CK chondrites: Implications for the metamorphism of the CK chondrite parent body. *Proceedings of the NIPR Symposium on Antarctic Meteorites* 6:204-233.

Norton M. B., and McSween H. Y., Jr. 2007. Quantifying the volumetric abundances of the components of the Murray CM chondrite: a preliminary investigation (abstract). *38. Lunar and Planetary Science Conference* #1807.

Pack A., Shelley M. G., and Palme H. 2004. Chondrules with peculiar REE patterns: implications for solar nebular condensation at high C/O. *Science* 303:997-1000.

Rubin A.E., 1984. Coarse-grained chondrule rims in type 3 chondrites. *Geochimica et Cosmochimica Acta* 48:1779–1789.

Rubin A. E. 1998. Correlated petrologic and geochemical characteristics of CO3 chondrites. *Meteoritics & Planetary Science* 33:385-391.

Rubin A.E., and Wasson J.T. 1987. Chondrules, matrix and coarse-grained chondrule rims in the Allende meteorite - origin, interrelationships, and possible precursor components. *Geochimica et Cosmochimica Acta* 51:1923–1937.

Rubin A. E., James J. A., Keck B. D., Weeks K. S., Sears D. W. G., and Jarosewich E. 1985. The Colony meteorite and variations in CO3 chondrite properties. *Meteoritics* 20:175-196.

Russell S. S., Huss G. R., Fahey A. J., Greenwood R. C., Hutchison R., Wasserburg G. 1998. An isotopic and petrologic study of calcium-aluminium inclusion from CO3 meteorites. *Geochimica et Cosmochimica Acta* 62:689-714.

Scott E. R. D. 2007. Chondrites and the Protoplanetary Disk. *Annual Review of Earth and Planetary Sciences* 35:577-620.

Scott E. R. D., and Krot A. N. 2006. Chondritic meteorites and the high-temperature nebular origins of their components. In: *Chondrites and the Protoplanetary Disk*. Krot A. N., Scott E. R. D. and Reipurth B. (eds.), ASP Conference Series, Vol. 341, San Francisco: Astronomical Society of the Pacific, 1029 p.

Scott E. R. D., Love S. G., and Krot A. N. 1996. Formation of chondrules and chondrites in the protoplanetary nebula. In: *Chondrules and the Protoplanetary disk*. Hewins R. H., Jones R. H., and Scott E. R. D. (es.). Cambridge University Press pp.87-96.

Simon S. B., and Haggerty S. 1979. Petrography and olivine mineral chemistry of chondrules and inclusions in the Allende meteorite (abstract). *10. Lunar and Planetary Science Conference* #1119- 1121.

Simon S. B., and Grossman L. 2004. A preferred method for the determination of bulk compositions of coarse-grained refractory inclusions and some implications of the results. *Geochimica et Cosmochimica Acta* 68:4237-4248.

Weber D. 1995. Refractory inclusions from the carbonaceous chondrite Acfer 094 (abstract). *Meteoritics* 30 #595.

Zipfel J., Wlotzka F., and Spettel B. 1998. Bulk chemistry and mineralogy of a new “unique” metal-rich chondritic breccia, Hammadah al Hamra 237 (abstract). *29. Lunar and Planetary Science Conference* #1417.

## Tables

Table 1: Reported modal abundances of refractory inclusions (Ca,Al-rich inclusions *and* amoeboid olivine aggregates).

| | ref. incl. [area%] |
|---|---:|
| CI | <0.01 |
| CM | 5 |
| CO | 13 |
| CV | 10 |
| CK | 4 |
| CR | 0.5 |
| CH | 0.1 |
| $Cb_a$ | <0.1 |
| $CB_b$ | <0.1 |
| | |
| H | 0.01-0.2 |
| L | <0.1 |
| LL | <0.1 |
| | |
| EH | <0.1 |
| EL | <0.1 |
| | |
| R | <0.1 |
| K | <0.1 |

ref. incl.: refractory inclusions. Data taken from Scott and Krot (2006).

Table 2: CAI modal abundances taken from various literature sources.

| **CV** | 1 | | 2 | | | | 3 | | | 4 | |
|---|---|---|---|---|---|---|---|---|---|---|---|
| | CAIs [area%] | points counted [No.] | CAIs [area%] | stud. area [$mm^2$] | wght. av. [area%] | av. CAI diam. [µm] | CAIs [area%] | size [mm] | stud. area [$mm^2$] | CAIs [area%] | stud. area [$mm^2$] |
| Vigarano | 5.3 | 1594 | 1.70 | 65 | 0.33 | | | | | | |
| Efremovka | 3.7 | 1611 | 1.67 | 63 | 0.32 | | | | | | |
| Mokoia | 3.5 | 1510 | 0.65 | 42 | 0.08 | | | | | | |
| Leoville | 6.6 | 1705 | 1.89 | 160 | 0.92 | 411 | | | | | |
| Kaba | 7.4 | 1561 | | | | | | | | | |
| Grosnaja | 2.8 | 1675 | | | | | | | | | |
| Bali | 4.0 | 1715 | | | | | | | | | |
| Arch | 2.8 | 1512 | | | | | | | | | |
| Coolidge | 2.5 | 1638 | | | | | | | | | |
| Allende | 9.4 | 1572 | | | | | 7.1 | up to >4 | 14 sect. | 2.52 | 17 sect. |
| mean | 4.8 | | 1.48 | | | | 7.1 | | | 2.52 | |
| s.d. | 2.3 | | 0.56 | | | | | | | | |
| total | | | | 330 | 1.65 | | | | | | 1900 |

| **CO** | 5 | | 2 | | | | 6 | | | 7 | |
|---|---|---|---|---|---|---|---|---|---|---|---|
| | CAIs [area%] | points counted [No.] | CAIs [area%] | stud. area [$mm^2$] | wght. av. [area%] | av. CAI diam. [µm] | CAIs [area%] | stud. area [$mm^2$] | wght. av. [area%] | CAIs [area%] | points counted [No.] |
| Warrenton (3.6) | 2.5 | 2852 | 0.85 | 67 | 0.31 | 187 | 0.97 | 67 | 0.07 | | |
| Lance (3.4) | 1.2 | 1533 | 1.38 | 50 | 0.38 | 200 | 0.98 | 50 | 0.05 | | |
| ALHA77307 (3.0) | | | 1.10 | 67 | 0.40 | 217 | 0.97 | 67 | 0.07 | 1.2 | 2250 |
| ALHA77003 (3.5) | | | | | | | 0.92 | 91 | 0.09 | | |
| ALH82101 (3.3) | | | | | | | 1.30 | 80 | 0.11 | | |
| Colony (3.0) | | | | | | | 0.99 | 112 | 0.12 | | |
| Colony 1 | | | | | | | | | | 1.0 | 1533 |
| Colony 2 | | | | | | | | | | 3.0 | 1000 |
| Colony 3 | | | | | | | | | | 3.6 | 1000 |
| Felix (3.2) | 3.5 | 1529 | | | | | 1.50 | 48 | 0.08 | | |
| Isna (3.7) | 1.6 | 3196 | | | | | 1.40 | 108 | 0.16 | | |
| Kainsaz (3.1) | 2.4 | 1765 | | | | | 0.66 | 128 | 0.09 | | |
| Ornans (3.3) | 1.4 | 2711 | | | | | 0.63 | 179 | 0.12 | | |
| CO3 | | | | | | | | | | 1.9(0.9-3.5) | |
| mean | 2.1 | | 1.11 | | | | 1.03 | | | 2.2 | |
| s.d. | 0.9 | | 0.27 | | | | 0.29 | | | 1.3 | |
| total | | 13586 | | 184 | 1.09 | | | 930 | 0.97 | | |

*Table 2 continued*

| **CM** | 8 | | |
|---|---|---|---|
| | CAIs [area%] | stud. area [mm²] | wght. av. [area%] |
| Murray 4b | 1.1 | 12.8 | 0.12 |
| Murray 6 | 3.8 | 12.8 | 0.42 |
| Murray 8a | 0.5 | 36.6 | 0.16 |
| Murray 8b | 1.2 | 16.3 | 0.17 |
| Murray 10 | 1.5 | 36.6 | 0.48 |
| | | | |
| mean | 1.6 | | |
| s.d. | 1.3 | | |
| total | | 115.2 | 1.35 |

| **CK** | | 9 | |
|---|---|---|---|
| | CAIs [area%] | stud. area [mm²] | wght. av. [area%] |
| Karoonda | 0.2 | ~220 | 0.2 |
| EET 87507 | 0.2 | ~140 | 0.2 |
| | | | |
| total | | ~360 | 0.2 |

| **CR** | 1 | |
|---|---|---|
| | CAIs [area%] | points counted [No.] |
| Renazzo | 0.3 | 1562 |
| Al Rais | 0.8 | 1684 |

| **CK/CV** | 10 | |
|---|---|---|
| | CAIs [area%] | size [µm] |
| Ningqiang | 1.0 | 200 |

| **ungr.** | 11 | |
|---|---|---|
| | CAIs [area%] | size [µm] |
| Acfer 094 | <2 | 40-500 |

Mean values are simple modal abundance averages. Weighted averages are calculated using eq. (6). A graphical representation of the CAI modal abundances is displayed in Fig. 1. wght. av.: weighted averages; stud. area: studied area; av. CAI diam.: average CAI diameter; size: CAI size; s.d.: standard deviation. Data taken from: [1]McSween (1977b); [2]May et al. (1999);

[3]Simon & Haggerty (1979); [4]Kornacki & Wood (1984); [5]McSween (1977a); [6]Russell et al. (1998); [7]Rubin et al. (1985) and references therein; [8]Norton & McSween (2007); [9]Noguchi (1993); [10]Kallemeyn et al. (1991); [11]Weber (1995)

Table 3: Edge lengths of single pixels in the X-ray maps used for CAI identification.

| | edge length of pixel [μm] |
|---|---|
| *CV* | |
| Allende 1 | 20 |
| Allende 3 | 14 |
| Allende 8 | 20 |
| *CO* | |
| Warrenton | 9 |
| Kainsaz | 24 |
| DaG 190 | 20 |
| *CM* | |
| Murchison | 10 |
| Nogoya | 9 |
| *CR* | |
| Renazzo | 13 |
| Acfer 209 | 20 |
| *CK* | |
| Karoonda | 19 |
| *CK/CV* | |
| DaG 055 | 20 |
| Ningqiang 1 | 14 |
| Ningqiang 2 | 16 |
| *CB* | |
| Isheyevo 06 | 18 |
| Isheyevo 2 | 21 |
| *ungr* | |
| Acfer 094 | 4 |

Table 4: CAI data of carbonaceous chondrites from this work.

| | CAIs [area%] | stud. area [mm²] | wght. av. [area%] | No of. CAIs | CAI model radius $r_m$ [µm] mean | min | max |
|---|---|---|---|---|---|---|---|
| *CV* | | | | | | | |
| Allende 1 | 5.94 | 220 | 2.08 | 223 | 140 | 27 | 1508 |
| Allende 3 | 4.65 | 282 | 2.08 | 495 | 92 | 18 | 869 |
| Allende 8 | 4.47 | 128 | 0.91 | 177 | 101 | 27 | 741 |
| mean | 5.02 | | | | | | |
| s.d. | 0.80 | | | | | | |
| total | | 630 | 5.06 | | | | |
| *CO* | | | | | | | |
| Warrenton (3.6) | 1.01 | 42 | 0.11 | 112 | 34 | 5 | 145 |
| Kainsaz (3.1) | 0.47 | 142 | 0.18 | 31 | 83 | 13 | 220 |
| DaG 190 | 1.35 | 197 | 0.70 | 208 | 64 | 22 | 331 |
| mean | 0.94 | | | | | | |
| s.d. | 0.44 | | | | | | |
| total | | 380 | 0.98 | | | | |
| *CM* | | | | | | | |
| Murchison | 0.97 | 34 | 0.82 | 201 | 23 | 6 | 180 |
| Nogoya | 0.02 | 6 | 0.002 | | 6 | 75 | 13 |
| mean | 0.49 | | | | | | |
| s.d. | 0.67 | | | | | | |
| total | | 41 | 0.83 | | | | |
| *CR* | | | | | | | |
| Renazzo | 0.09 | 112 | 0.04 | 60 | 24 | 7 | 57 |
| Acfer 209 | 0.14 | 155 | 0.08 | 31 | 48 | 22 | 143 |
| mean | 0.12 | | | | | | |
| s.d. | 0.04 | | | | | | |
| total | | 267 | 0.12 | | | | |
| *CK* | | | | | | | |
| Karoonda | 5.74 | 146 | | 665 | 63 | 15 | 313 |
| *CK/CV* | | | | | | | |
| DaG 055 | 0.59 | 169 | 0.28 | 76 | 65 | 11 | 226 |
| Ningqiang 1 | 0.38 | 62 | 0.07 | 10 | 87 | 8 | 189 |
| Ningqiang 2 | 1.22 | 123 | 0.42 | 36 | 115 | 9 | 315 |
| mean | 0.73 | | | | | | |
| s.d. | 0.44 | | | | | | |
| total | | 354 | 0.77 | | | | |
| *CB* | | | | | | | |
| Isheyevo 06 | 0.11 | 234 | 0.08 | 62 | 36 | 14 | 101 |
| Isheyevo 2 | 0.06 | 101 | 0.02 | 32 | 24 | 12 | 53 |
| mean | 0.09 | | | | | | |
| s.d. | 0.04 | | | | | | |
| total | | 335 | 0.09 | | | | |
| *ungr.* | | | | | | | |
| Acfer 094 | 1.12 | 22 | | 298 | 16 | 3 | 59 |

wght. av.: weighted averages; stud. area: studied area.

Table 5: Calculated CAI modal abundances using bulk chondrite Al concentrations.

| | 1 | 2 | 3 | 4 | 5 | 6 | 7 | 8 | 9 | 10 | 11 | 12 | 13 | 14 | 15 |
|---|---|---|---|---|---|---|---|---|---|---|---|---|---|---|---|
| | bulk chondrite concentrations [wt%] | | | | $Al_u$ [wt%] | | | $\Delta Al = Al_{chondrite} - Al_u$ [wt%] | | | CAIs [area%] | | | | max. CAIs |
| | Al | Mg | Si | Fe | x = Mg | x = Si | x = Fe | x = Mg | x = Si | x = Fe | x = Mg | x = Si | x = Fe | this study | [area%] |
| CI | 0.87 | 9.70 | 10.64 | 18.20 | 0.87 | 0.87 | 0.87 | 0.00 | 0.00 | 0.00 | 0.00 | 0.00 | 0.00 | 0.00 | 4.81 |
| CM | 1.13 | 11.50 | 12.70 | 21.30 | 1.03 | 1.03 | 1.01 | 0.10 | 0.10 | 0.12 | 0.58 | 0.54 | 0.65 | 1.21 | 6.28 |
| CV | 1.68 | 14.30 | 15.70 | 23.50 | 1.28 | 1.28 | 1.12 | 0.40 | 0.40 | 0.56 | 2.25 | 2.24 | 3.13 | 2.98 | 9.33 |
| CO | 1.40 | 14.50 | 15.80 | 25.00 | 1.29 | 1.28 | 1.19 | 0.11 | 0.12 | 0.21 | 0.59 | 0.64 | 1.18 | 0.99 | 7.78 |
| CK | 1.47 | 14.70 | 15.80 | 23.00 | 1.31 | 1.28 | 1.09 | 0.16 | 0.19 | 0.38 | 0.88 | 1.03 | 2.09 | 0.20 | 8.17 |
| CR | 1.15 | 13.70 | 15.00 | 23.80 | 1.22 | 1.22 | 1.13 | -0.07 | -0.07 | 0.02 | - | - | 0.10 | 0.12 | 6.39 |
| CH | 1.05 | 11.30 | 13.50 | 38.00 | 1.01 | 1.10 | 1.81 | 0.04 | -0.05 | -0.76 | 0.24 | - | - | <0.1 | 5.83 |
| CB[1] | 0.39 | 3.91 | - | 70.95 | 0.35 | - | 3.37 | 0.04 | - | -2.98 | 0.23 | - | - | 0.10 | 2.17 |
| H | 1.06 | 14.10 | 17.10 | 27.20 | 1.26 | 1.39 | 1.29 | -0.20 | -0.33 | -0.23 | - | - | - | <0.1 | 5.89 |
| L | 1.16 | 14.90 | 18.60 | 21.75 | 1.33 | 1.51 | 1.03 | -0.17 | -0.35 | 0.13 | - | - | 0.70 | <0.1 | 6.44 |
| LL | 1.18 | 15.30 | 18.90 | 19.80 | 1.36 | 1.54 | 0.94 | -0.18 | -0.36 | 0.24 | - | - | 1.33 | <0.1 | 6.56 |
| EH | 0.82 | 10.73 | 16.60 | 30.50 | 0.96 | 1.35 | 1.45 | -0.14 | -0.53 | -0.63 | - | - | - | <0.1 | 4.56 |
| EL | 1.00 | 13.75 | 18.80 | 24.80 | 1.23 | 1.53 | 1.18 | -0.23 | -0.53 | -0.18 | - | - | - | <0.1 | 5.56 |
| R | 1.06 | 12.90 | 18.00 | 24.40 | 1.15 | 1.46 | 1.16 | -0.09 | -0.40 | -0.10 | - | - | - | <0.1 | 5.89 |
| K | 1.30 | 15.40 | 16.90 | 24.70 | 1.37 | 1.37 | 1.17 | -0.07 | -0.07 | 0.13 | - | - | 0.70 | <0.1 | 7.22 |

Columns 1-4: reported bulk chondrite element-concentrations (Lodders and Fegley, 1998; [1]Zipfel et al. 1998). Columns 5-7: Calculated, unchanged Al concentrations ($Al_u$) using eq. (4) in the chondrites as if these had not been changed relative to the CI-chondritic value and under the assumption that the element $x$ (= $x_{chondrite}$ in eq. 4) also had not changed. Columns 8-10: ΔAl is the difference between the measured bulk chondrite Al concentrations (column 1) and $Al_u$ (columns 5-7) for different $x$. Columns 11-13: Calculated CAI modal abundances using ΔAl and eq. (5) for different $x$. Column 14: CAI modal abundances that we determined in this study (cf. Table 6). Column 15: Maximum CAI modal abundances calculated from the bulk chondrite Al concentrations and assuming that all Al was delivered by CAIs.

Table 6: CAI modal abundances deduced from our and previous studies.

| | CAIs [area%] | error [2$\sigma$] | area [$mm^2$] |
|---|---|---|---|
| CI | <0.01 | | - |
| CM | **1.21** | $^{+1.0}_{-0.5}$ | 156 |
| CO | **0.99** | $^{+1.0}_{-0.3}$ | 1494 |
| CV | **2.98** | $^{+0.3}_{-0.1}$ | 2860 |
| CK | **0.2** | $^{+0.2}_{-0.2}$ | 506 |
| CK/CV[1] | **0.77** | $^{+0.9}_{-0.4}$ | 354 |
| CR | **0.12** | $^{+0.15}_{-0.1}$ | 267 |
| CH | 0.1 | | - |
| $CB_a$ | <0.1 | | - |
| $CB_b$ | **0.1** | $^{+0.15}_{-0.1}$ | 335 |
| Acfer 094 | **1.12** | $^{+1.5}_{-0.7}$ | 22 |
| H | 0.01-0.2 | | - |
| L | <0.1 | | - |
| LL | <0.1 | | - |
| EH | <0.1 | | - |
| EL | <0.1 | | - |
| R | <0.1 | | - |
| K | <0.1 | | - |

In bold our new CAI modal abundances. CAI modal abundances have been calculated as weighted averages using data from: CM - Norton & McSween (2007), this work; CO - Russell et al. (1998), May et al. (1999), this work; CV - Kornacki & Wood (1984), May et al. (1999), this work; CK - this work; CK/CV - Kallemeyn et al. (1991); CR, CB, Acfer 094 - this work. estim. error: estimated error; area: area used to determine the CAI abundance. [1]These are: Ningqiang & DaG 055.

**Figures**

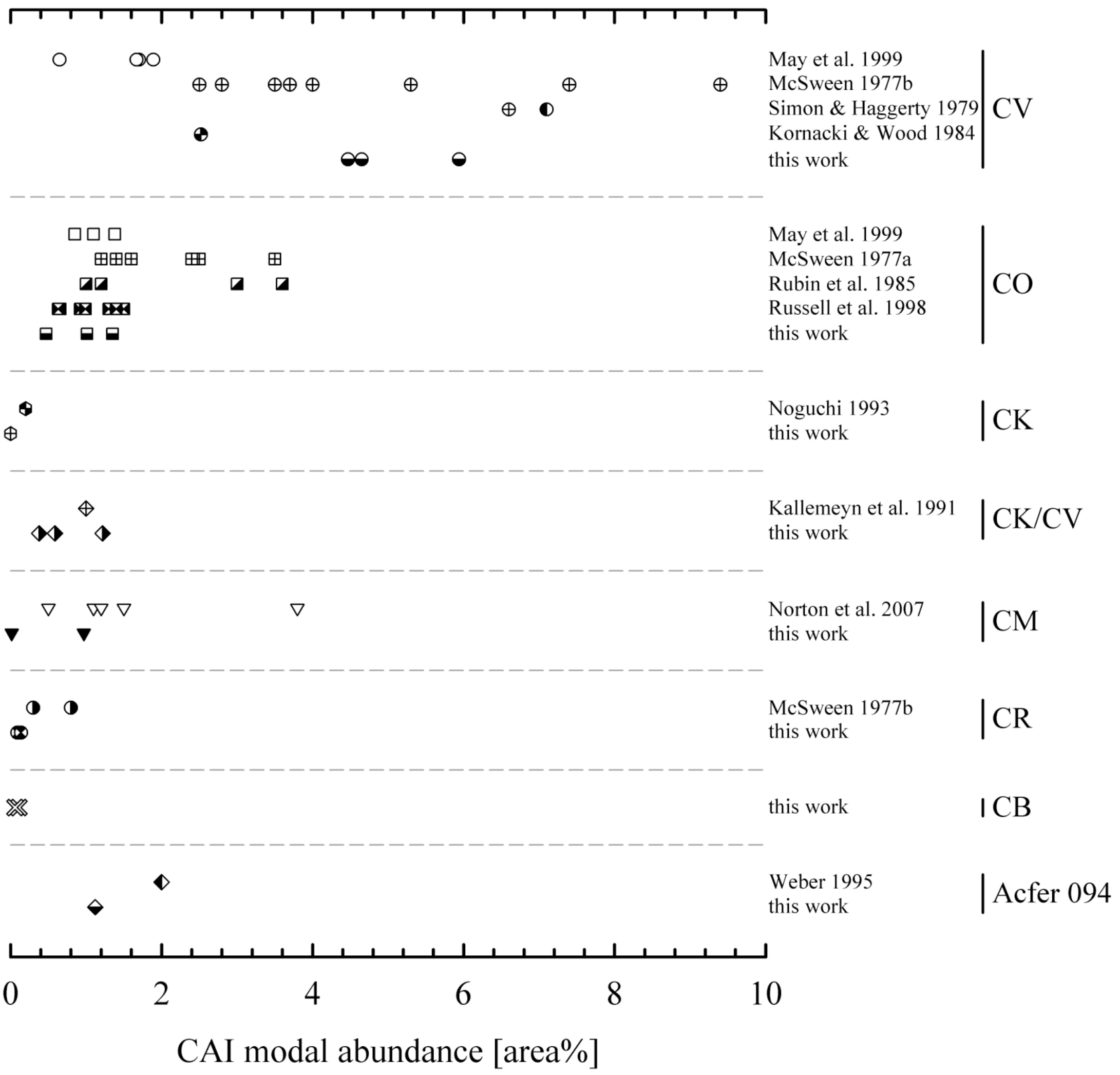

Fig. 1

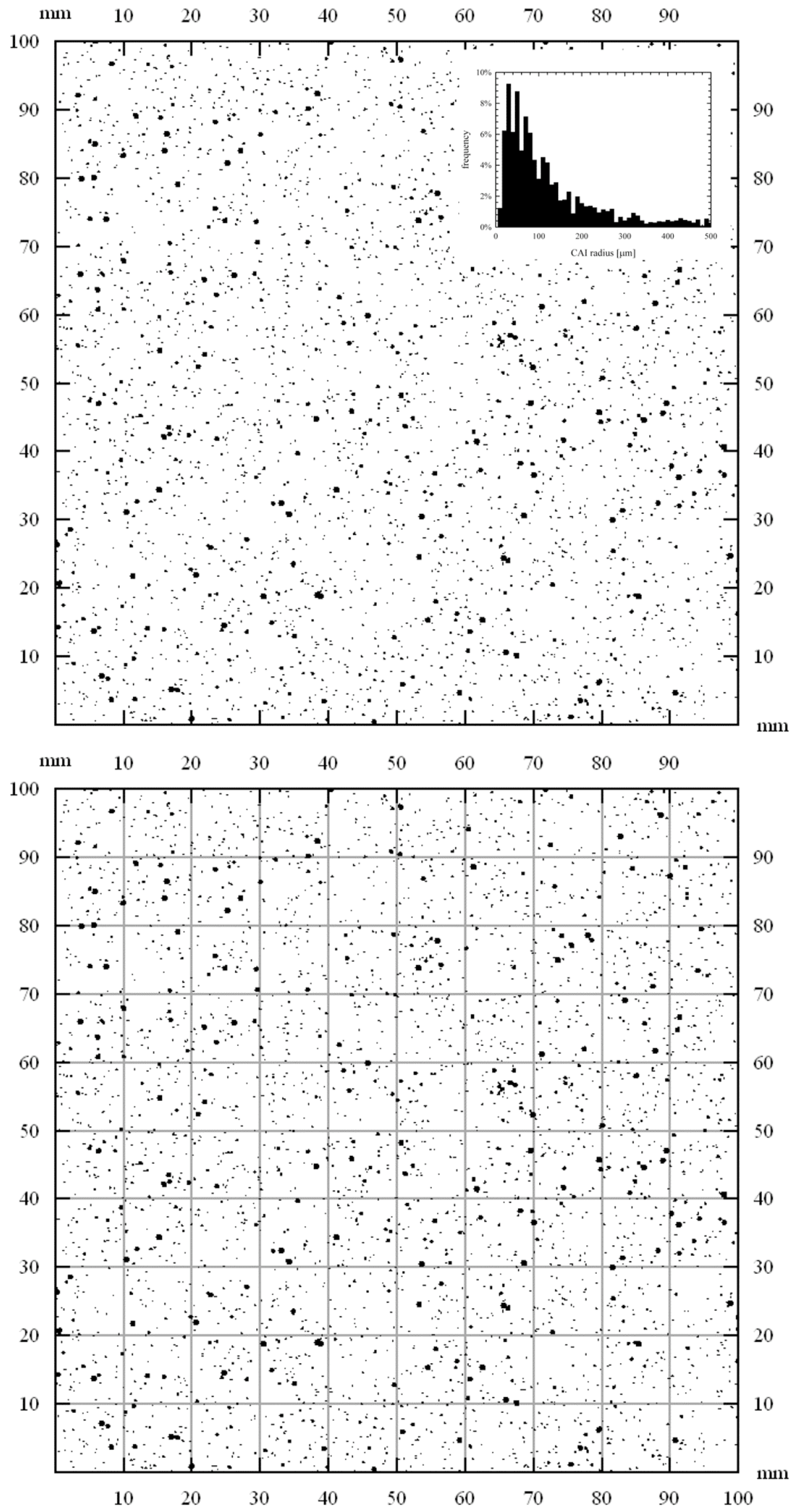
mm
10 20 30 40 50 60 70 80 90 100
frequency
0% 2% 4% 6% 8% 10%
0 100 200 300 400 500
CAI radius [µm]

Fig. 2

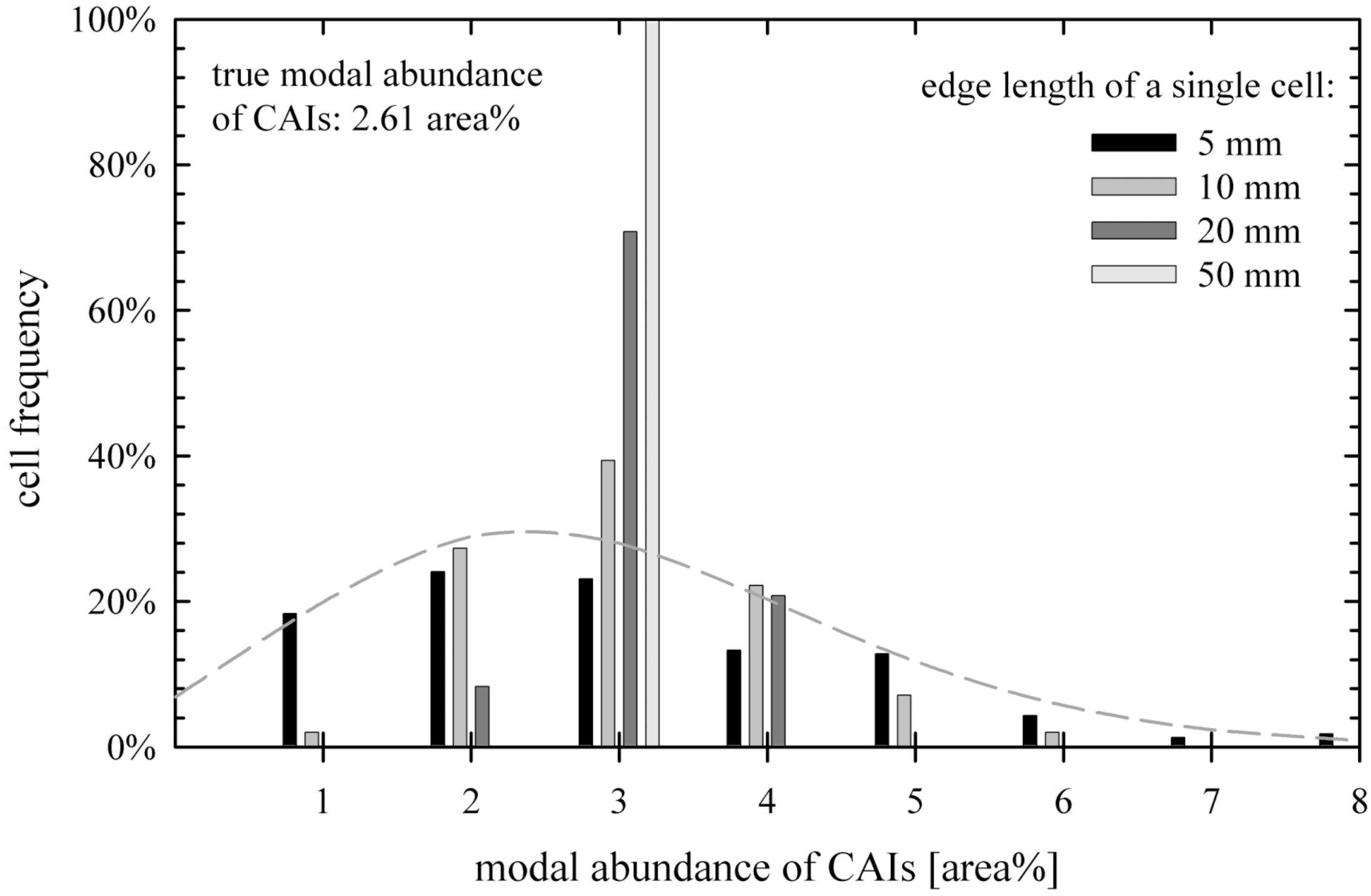

Fig. 3

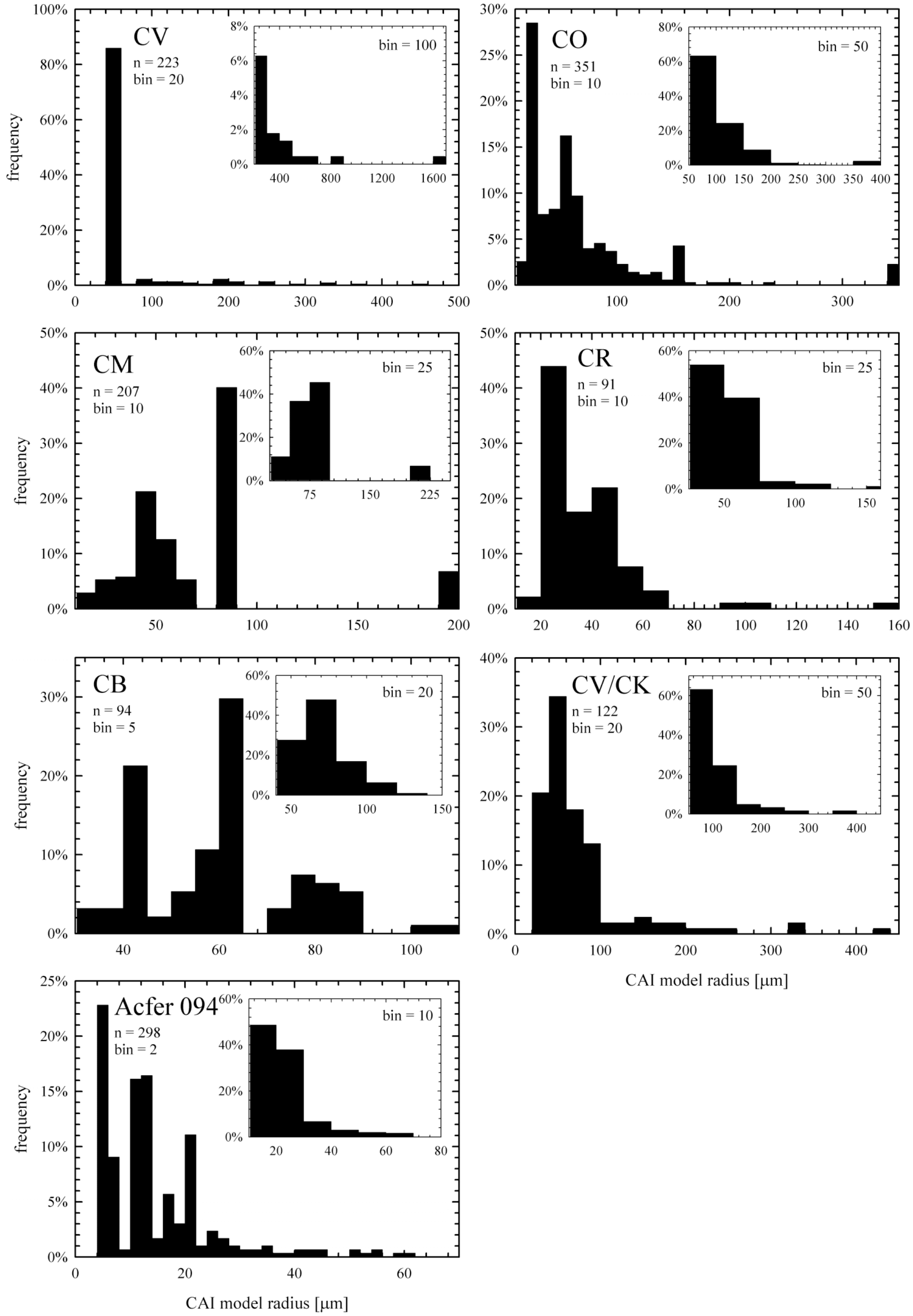

Fig. 4

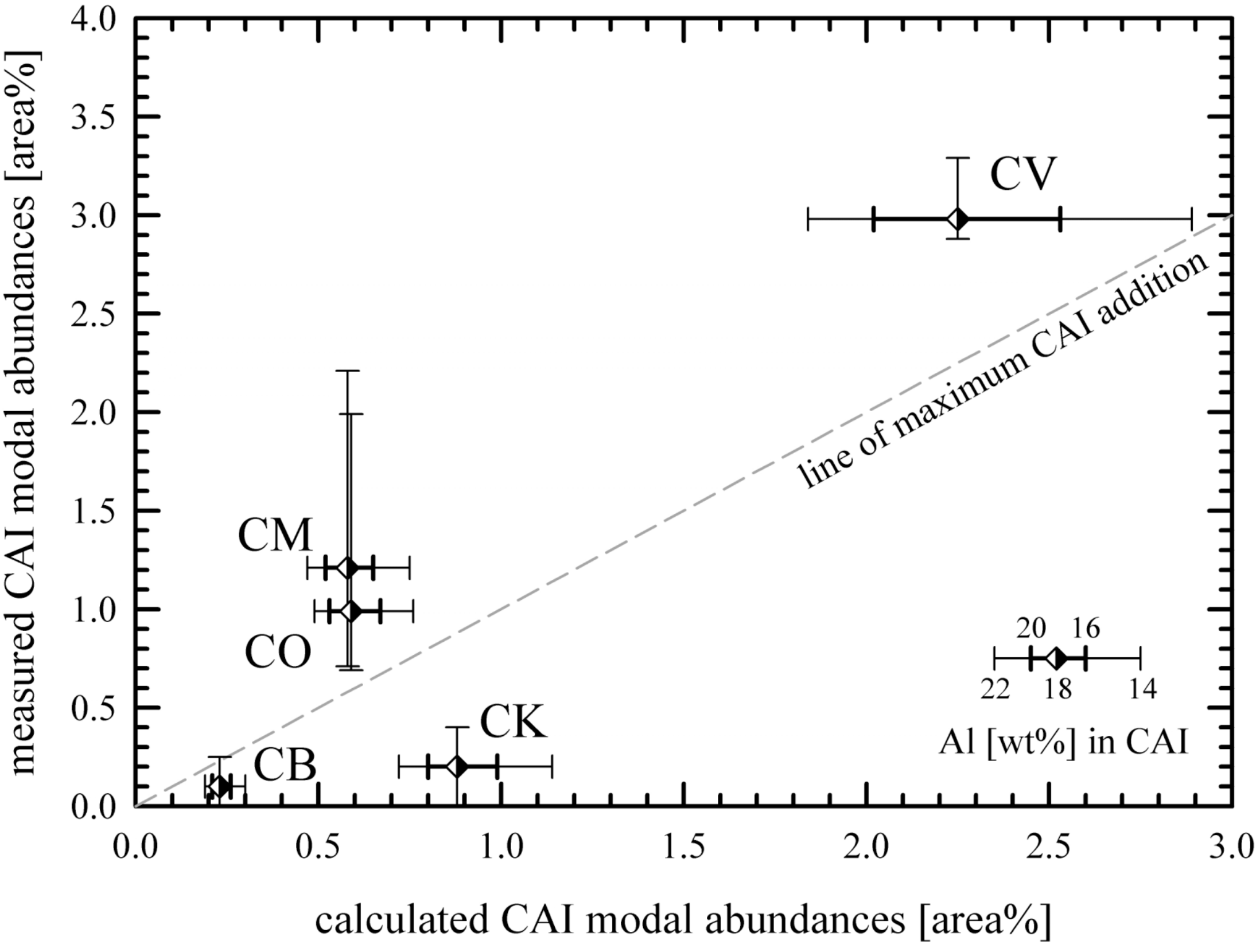

Fig. 5

## Figure captions

Fig. 1: Compilation of all literature data of CAI modal abundances and this work for carbonaceous chondrites (see Table 2). A Poisson distribution can be recognised in the scatter of CAI abundance of individual chondrite groups: data points are concentrated a bit to the left of the average of the smallest and the largest modal abundance and thin out towards the end members. The available data are, however, insufficient to produce reasonable histograms to show this.

Fig. 2: Screenshots of the computer model simulating the CAI distribution in a meteorite. The CAIs (black dots) have been randomly placed in this 100x100mm sized square. CAI sizes follow a log-normal distribution with a maximum radius of 500 µm resulting in an average radius of 113 µm. The inset displays the CAI size distribution used for this figure and Fig. 3. The modal abundance is 2.61 area%. The grid in (b) illustrates the non-homogeneous distribution of CAIs, which in fact follow a Poisson distribution and demonstrates why some thin sections are likely to have much more or much fewer CAIs than average.

Fig. 3: Histograms produced using the distribution displayed in Fig. 2. The four different histograms are produced using different grid spacings, designated 'single cell edge length'. The smaller the spacing the more the peak of the distribution shifts to lower values. The grey dashed lined indicates a Poisson distribution for a spacing of 10 mm.

Fig. 4: CAI size distributions obtained from our measurements. Two bin ranges are given for all chondrite groups to account for possible artefacts when choosing bin ranges.

Fig. 5: Calculated versus measured CAI modal abundances for chondrites with excess Al (cf. Table 5). Chondrites plot on the 'line of maximum CAI addition' if the Al excess in a chondrite equals the Al contributed by CAIs. If chondrites plot above the line, their initial bulk Al concentration must have been lower than the CI-chondritic value and vice versa. Two

horizontal error bars represent calculated CAI modal abundances using 14 and 22 (thin bar) and 16 and 20 (bold bar) wt% Al in bulk CAIs. Vertical error bars are those listed in Table 6.

## Electronic Appendix

EA 1: False colour X-ray image of the Allende1 section. Al (white), Ca (yellow), Mg (green), Si (blue) and Fe (red). Some CAIs are indicated by arrows.

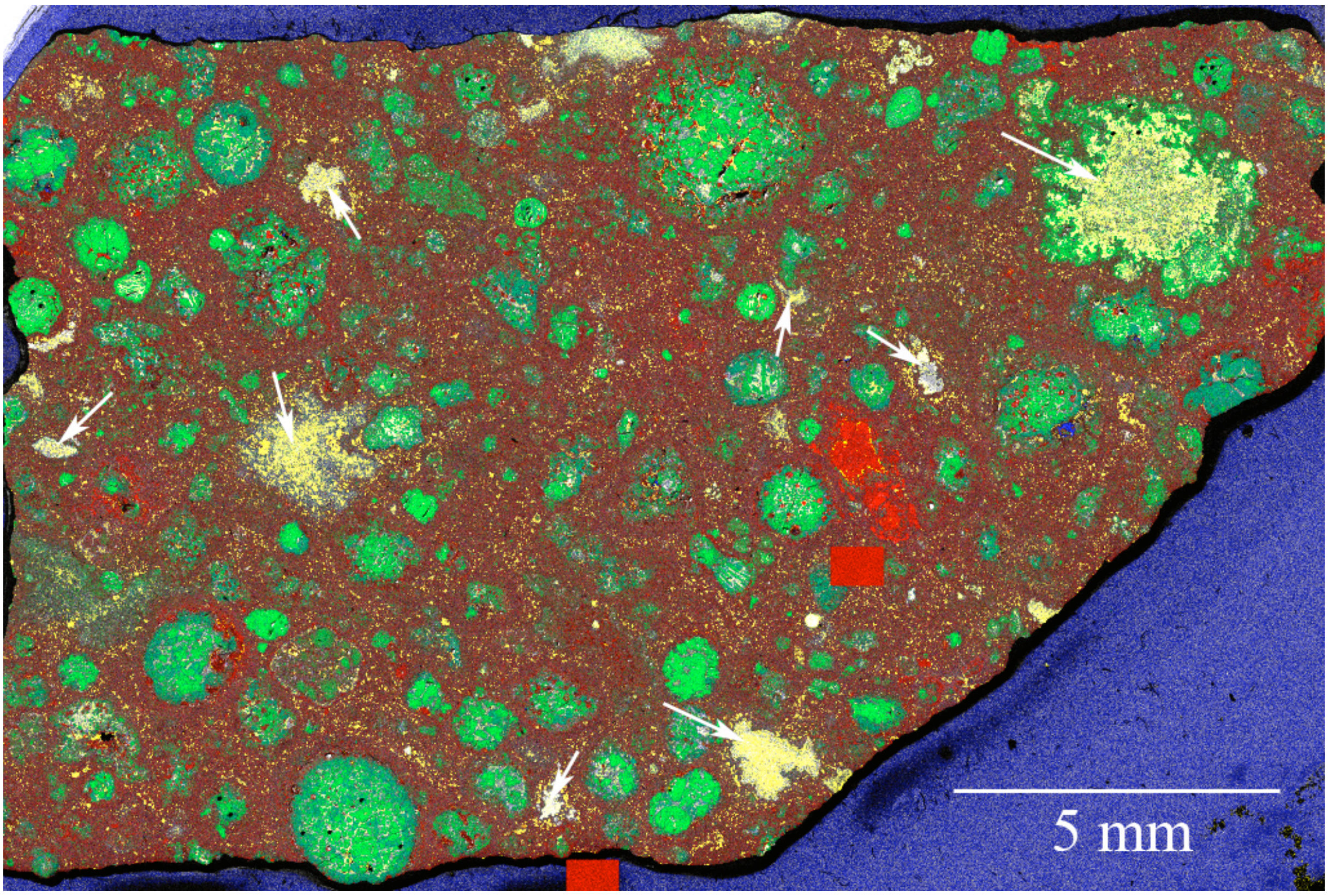